\documentstyle[epsfig,11pt]{article}

\def\be{\begin{eqnarray}}
\def\ee{\end{eqnarray}}
\def\dbmu{\partial_{\mu}}
\def\dhmu{\partial^{\mu}}

\def\dhsigma{\partial^{\sigma}}
\def\ie{{\it i.e. }}

\def\Imm{\Im m}
\def\partialleftright{\stackrel{\leftrightarrow}{\partial^\delta}}

\begin{document}
\begin{titlepage}

\begin{flushright}
CERN-TH/97-79\\
LPTHE-ORSAY 97/17 \\
Saclay T97/038 \\
hep-th/9704177 \\
\end{flushright}

\vskip.5cm
\begin{center}
{\huge{\bf Four-dimensional $M$-theory and supersymmetry breaking}}
\end{center}
\vskip1.5cm

\centerline{ Emilian Dudas $^{a,b}$ {\it and}\footnote{e-mail addresses:
 edudas@mail.cern.ch 
{\it and} grojean@spht.saclay.cea.fr} Christophe Grojean $^{a,c}$} 
\vskip 15pt
\centerline{$^{a}$ CERN-TH}
\centerline{CH-1211 Geneva 23, {\sc Switzerland}}
\vskip 3pt
\centerline{$^{b}$ Laboratoire de Physique Th\'eorique et Hautes Energies}
\centerline{B\^at. 211, Univ. Paris-Sud, F-91405 Orsay Cedex, {\sc France}}
\vskip 3pt
\centerline{$^{c}$ CEA-SACLAY, Service de Physique Th\'eorique}
\centerline{F-91191 Gif-sur-Yvette Cedex, {\sc France}}
\vglue .5truecm

\begin{abstract}
\vskip 3pt
We investigate compactifications of $M$-theory from $11\to 5\to 4$
dimensions and discuss 
geometrical properties of 4-d moduli fields related to the structure of 
5-d theory.

We study supersymmetry breaking by compactification of the fifth 
dimension and find that an universal superpotential is generated for the 
axion-dilaton 
superfield $S$.
The resulting theory has a vacuum with $<S>=1$, zero cosmological constant 
and a gravitino mass depending on the fifth radius as
$m_{3/2} \sim R_5^{-2}/M_{Pl}$.  

We discuss phenomenological aspects of this scenario, mainly the string 
unification and the decompactification problem.
\end{abstract}

\vfill
\begin{flushleft}
CERN-TH/97-79\\
April 1997\\
\end{flushleft}

\end{titlepage} 
\section{Introduction}
In trying to describe four-dimensional physics from compactified 
string theories, it has soon appeared \cite{stringscale85} that the 
ten-dimensional string 
can hardly be weakly coupled, leading to a far too large Newton's constant.
So phenomenological attention must be paid to strongly coupled strings.
Recently \cite{Mtheory}-\cite{Horava&witten}, many progresses have been made 
in understanding 
this new physics: the strongly coupled regime is now viewed as the low energy 
limit of 
$M$-theory and in particular, the strongly coupled 
$E_8 \times E_8$ heterotic string, traditionally considered as the most 
relevant one for 
phenomenology, can be
described, in the low energy limit, by the eleven-dimensional supergravity
with the two $E_8$ gauge factors living each on a 10-d boundary.
The radius of the eleventh dimension is related to the string coupling by 
$R_{11} \sim 
\lambda_{st}^{2/3}$. 
So in the strongly coupled regime,
$R_{11}$ has to be large, in particular, could be larger than the typical 
radius of the
 six other compact dimensions.
Therefore, it appears in that case that the eleventh dimension has to be 
compactified 
after the Calabi-Yau internal manifold 
(\cite{phenoMtheory},\cite{aq},\cite{li}).
Thus describing four-dimensional physics from $E_8 \times E_8$ heterotic 
strongly 
coupled 
string should be equivalent to compactify the eleven-dimensional 
supergravity on a 
Calabi-Yau 
and then compactify the fifth dimension on $S^1/Z_2$.
Our goal is  to compactify the Lagrangian to 4-d in a way compatible
with N=1 4-d supersymmetry. As shown in \cite{Horava&witten}, the 
presence of
the boundaries and the interaction between the boundary fields and the
bulk fields make this task difficult, as the 7-d internal space
is not really a direct product $Q \times  S^1 /Z_2$. However, we shall be
mainly concerned in compactifying the (bulk) gravitational sector of the 
theory, 
where this 
difficulty do not appear and simply add the kinetic terms
for the gauge fields on the boundaries. We ignore all the matter fields
in 4-d and their interactions.
This is the point of view adopted here and we  compare this pattern of 
compactification with 
the previous one studied in \cite{Dudas&Mourad} and which corresponds to 
$11\to 10 \to 4$.

Performing an explicit compactification on a CY manifold can be rather 
difficult, 
so we adopt 
an alternative way, by truncating with a symmetry 
of the compact space such that to maintain $N=1$ supersymmetry
in 4-d. Of course, in this case we can describe only the analog of
untwisted fields of string theories, originating from 11-d and 10-d
fields.

In the section 2 of the paper, we identify, for different projections, the 
K\"ahler
 structure for 
the moduli fields describing the shape of the  internal manifold. We will 
observe 
very interesting 
geometrical properties, namely
in all the cases the size of the compactified manifold is contained
exclusively in the dilaton-axion superfield $S$, all the other moduli
fields being invariant under dilatations of the 6-d compactified space.

Our main goal, to be studied in section 3, is the $N=1$ spontaneous 
supersymmetry
 breaking in four 
dimensions by compactification from 5-d to 4-d, by using the 
Scherk-Schwarz 
mechanism 
\cite{Scherk&Schwarz}. We argue that we obtain
results that look like non-perturbative from the perturbative heterotic
string point of view, like an universal superpotential generation for $S$.
The corresponding model spontaneously breaks supersymmetry with a zero
cosmological constant, has the invariance $S \rightarrow 1/S$ and the minimum 
is reached for $S=1$.

 Section 4 shows that the use of the eleventh (or fifth, after the CY
compactification) dimension to break supersymmetry offers a new
perspective on the decompactification problem and the unification
problem of perturbative string theories. The dependence of the gravitino mass
on the fifth radius $R_5$ (seen from 5-d in Einstein units) is
\be
m_{3/2} \sim {{R_5^{-2}}\over{M_{Pl}^{(4)}}} \ . 
\ee
By using this result, we argue that a value of the
fifth radius of the order
$10^{12} GeV$ could solve both above-mentioned problems.

 We end with some conclusions and prospects.

\section{Dimensional reduction}

The bosonic part of the strongly coupled $E_8 \times  E_8$ heterotic
string Lagrangian is (we neglect for the moment higher-derivative terms) 
\cite{Horava&witten}
\be
&&S^{(11)}={1\over \kappa_{11}^2}\int_{M^{11}}d^{11}x\, \sqrt{g}
\left(
-{1\over 2} {\cal R}^{(11)} 
-{1\over 48}G_{IJKL}G^{IJKL}  
\right) 
\label{actionsugra11d}\\
&& -{\sqrt 2 \over 3456 \kappa_{11}^2} \int_{M^{11}} d^{11}x\,
\epsilon^{I_1\dots I_{11}} C_{I_1I_2I_3} G_{I_4\dots I_7} 
G_{I_8\dots I_{11}} 
- {1 \over {8 \pi (4 \pi \kappa_{11}^2 )^{2/3}}} 
\int_{M^{10}} d^{10}x
\sqrt{g}  tr F_{AB} F^{AB} \ , 
\nonumber  
\ee
where $I,J,K,L = 1 \dots 11$ and $G_{IJKL}$ is related to the field-strength 
of the three-form $C_{IJK}$ by (in differential form language)
\be
G = 6 d C + 
{{\kappa_{11}^{2/3} \delta (x^{11}) } \over {2 \sqrt{2} \pi (4 \pi )^{2/3} }} 
 dx^{11}  \omega_{3Y}
\ , 
\label{defG}
\ee
which is similar to the ten-dimensional relation $H=d B -{{\alpha'} \over 2}
\omega_{3Y}$.  
In (\ref{actionsugra11d}), $\kappa_{11}$ is the 11-d 
gravitational coupling which 
defines the 11-d 
Newton constant $\kappa_{11}=M_{11}^{-2/9}$, which is the $M$-theory scale.
For the purposes of this section we set $M_{11}=1$. The mass units will be
discussed in detail later on.

The Lagrangian above must be supplemented  by the Horawa-Witten 
$Z_2$ projection, 
which project out
 one gravitino and part of the other
fields on the boundary. It acts in the following way:
\be
x_{11} \rightarrow - x_{11}, \
\Psi_I (-x_{11}) = \Gamma_{11} \Psi_I (x_{11}) , 
\label{actionofHW}
\ee
where  $\Psi_I$ is the 11-d gravitino and $\Gamma_{11}=\Gamma_1 \cdots
\Gamma_{10}$ is the 10-dim chirality matrix, while the three form $C$ is 
odd and the
 metric tensor is even.

Throughout the paper,
$x^6,\cdots,x^{11}$ will denote coordinates on the CY manifold $Q$, 
$x^5$, the extra eleventh dimension, and $x^1,\cdots,x^4$, the ordinary 
four-dimension space-time.
We introduce a complex structure on $Q$ by defining
\be
y^i = {{x^{2i+4}+i\, x^{2i+5}}\over{\sqrt{2}}} , \ \ \ 
y^{\bar{\imath}} = {{x^{2i+4}-i\, x^{2i+5}}\over{\sqrt{2}}}, \ \ \ 
i=1,2,3.
\ee
In the following,
$\mu,\nu,\cdots$ will refer to 4-d and 5-d Lorentz indices (to be 
distinguished whenever
necessary) and $i,j,k \cdots$ to compact indices.

We truncate the Lagrangian (\ref{actionsugra11d}) in order to obtain N=2 
models\footnote{We call N=2 supersymmetry the smallest possible
supersymmetry in 5-d.}
in 5-d in the supergravity (Einstein) units, which we argue to be the
natural units in $M$-theory. Then we impose the Horava-Witten projection
 (which acts by
substituting $x_{11} \rightarrow x_{5}$ and $\Gamma_{11} \rightarrow 
\Gamma_{5}$ in (\ref{actionofHW})), which gives N=1 models in 4-d.

A simple dimensional reduction can be performed to obtain an action in 
five dimensions. Our way of truncation (see below) is such that,
 in the metric tensor, there is no mixing between compact and non-compact 
indices. So, going into supergravity coordinates in 5-d, we take
\be
g^{(11)}_{\mu\nu} = G^{-1/3} g^{(5)}_{\mu\nu}\, , \ \
g^{(11)}_{ij} = g_{ij}\ ,
\ee
where $G$ is the determinant of the metric in the compact space $g$ (to be distinguished form $g^{(5)}$).

Similarly, the relevant components of the three form are
\be
C_{\mu\nu\rho}, \ C_{\mu i {\bar \jmath}}= A_{\mu}^{(i {\bar \jmath})}, \
C_{ijk}=a \epsilon_{ijk}, \  
C_{i {\bar \jmath} {\bar k}} \ ,
\ee
where $a$ and $C_{i {\bar \jmath} {\bar k}}$ are complex scalars. 
Moreover we shall put $C_{i {\bar \jmath} {\bar k}}=0$ in the following, 
these fields being irrelevant for our analysis.

After algebraic manipulations, we obtain the desired 5-d bosonic 
action (see \cite{cadavid})
\be
&&
S^{(5)} = {-1\over  {2 }}
\int d^5x \sqrt{g^{(5)}}
\left[
{\cal R}^{(5)} + {1\over{12}} tr ( g^{-1} \dbmu g) tr ( g^{-1} \dhmu g)
+{1\over 4} tr ( g^{-1} \dbmu g g^{-1} \dhmu g) 
\right.\nonumber\\
&&\
+{1\over 24} G\, G_{\mu\nu\rho\sigma} G^{\mu\nu\rho\sigma}
+18 \, G^{1/3} (g^{i k }g^{{\bar \jmath}{\bar l}} - g^{i {\bar l}} 
g^{k {\bar \jmath}} ) 
F_{\mu\nu}^{(i {\bar \jmath})} F^{\mu\nu\, (k {\bar l})} 
\nonumber\\
&&\
\left.
+36 \left( \det g^{ij} \,  (\dbmu a)(\dhmu a) +
2 \det g^{i {\bar \jmath}} (\dbmu a)(\dhmu a^\dagger) +
\det g^{{\bar \imath} {\bar \jmath}} 
(\dbmu a^\dagger)(\dhmu a^\dagger) \right) 
\right] \nonumber\\
&&\
-{36 \sqrt{2}}
\int d^5x\,  
\epsilon^{\mu\nu\rho\sigma\tau}
\left( i\, C_{\mu\nu\rho} (\partial_\sigma a) (\partial_\tau a^\dagger)
+{1\over 4} \epsilon_{ikm} \epsilon_{{\bar \jmath} {\bar l} {\bar m}}
A_{\mu}^{(i {\bar \jmath})} F_{\nu\rho}^{(k {\bar l})} 
F_{\sigma\tau}^{(m {\bar n})}
\right) 
\nonumber\\
&& \ 
-{1\over {2\pi (4\pi)^{2/3}}} \int d^4x \, \sqrt{g^{(4)}} \, G^{1/2} 
tr F_{AB}F^{AB}
\ .
\ee
In the following, we perform the rescalings $(a,\, 
A_{\mu}^{(i {\bar \jmath})})
 \to (a,\, A_{\mu}^{(i {\bar \jmath})}) /6$.
We now consider particular truncations and also compactify the fifth 
dimension to obtain an 4-d 
action. In each case we discuss, the gauge group on the observable boundary  
is broken by the embedding of the spin connection into the gauge group.

To begin with, consider the simplest truncation, corresponding to a
compactified space with just one radius $e^{\sigma /2}$, the "breathing"
 mode.\footnote{This simple truncation in $M$-theory regime has been 
recently performed also in \cite{li}.}
Following Witten \cite{Witten85}, we pick up an $SU(3)$ subgroup of the 
$SU(4)\simeq SO(6)$ rotational group acting on $x^6,\cdots,x^{11}$. 
We define this $SU(3)$ such that $(y^1,y^2,y^3)$ transforms like a 
representation 3 and 
$(y^{\bar 1},y^{\bar 2},y^{\bar 3})$, like a representation ${\bar 3}$.
The 11-d fields have to respect this $SU(3)$ symmetry and also the $Z_2$ 
symmetry acting on $x^5$. The observable gauge group is $E_6$. 
The massless spectrum in 5-d is
the universal hypermultiplet $(C_{\mu\nu\rho} , e^{3 \sigma },
a,a^{\dagger})$ and the gravitational multiplet, corresponding to a
CY manifold with $h_{(1,1)} =1$ and  $h_{(2,1)} =0$.
The resulting 11-d tensor metric is written
\be
&g^{(11)}_{i {\bar \jmath}} = e^{\sigma} \delta_{i {\bar \jmath}}\, ,\ 
g^{(11)}_{\mu\nu} = e^{-2\sigma}  g^{(5) }_{\mu\nu}\, ,&\nonumber \\
&g^{(5)}_{55} = e^{2\gamma} \, , \  g^{(5)}_{\mu\nu} = e^{-\gamma}  
g^{(4) }_{\mu\nu}.& \label{metricto5d}
\ee
We have adopted here supergravity coordinates in which the 5-d and 4-d 
actions have a canonical supergravity expression. The scalar field 
$e^{\sigma}$ is related to the radius of the CY manifold, while 
$e^{\gamma}$ is related to the radius of the fifth 
dimension.
Similarly, in 4-d and after the Horava-Witten ($Z_2^{HW}$) projection,
 the three-form has only the massless components
\be
C^{(11)}_{\mu\nu 5} = C_{\mu\nu 5} ,\
C^{(11)}_{5 i {\bar \jmath}} = B \delta_{i {\bar \jmath}} ,\
C^{(11)}_{i j k } = \epsilon_{ijk} a,\
C^{(11)}_{{\bar \imath} {\bar \jmath} {\bar k} } = \epsilon_{ijk} 
a^\dagger \ .
\ee
$C_{\mu \nu 5}$ is a two-form in the 4-d space-time while $B$ is a
4-d scalar field.

The dimensional truncation of $S^{(5)}$ is easily performed and leads to
\be
&&S^{(4)} =
-\int d^{4}x \sqrt{g^{(4)}}
\left\{
{1\over 2} {\cal R}^{(4)} +
{3\over 4} (\dbmu \gamma) (\dhmu \gamma) +
{9 \over 4} (\dbmu \sigma) (\dhmu \sigma) 
\right.\label{action4-d}\\
&&
\ \ \left.+
{1\over 12}
e^{6\sigma} G_{\mu\nu\rho 5}G^{\mu\nu\rho 5} +
{3\over 4} e^{-2\gamma} (\dbmu B) (\dhmu B)
+{1 \over {2 \pi (4 \pi )^{2/3}}} e^{3\sigma} F_{\mu\nu} F^{\mu\nu} 
\right\} \ . \ 
\nonumber
\ee

To be consistent with a true CY compactification, this 4-d action has to 
derive from a K\"ahler potential; the supergravity coordinates are the 
appropriate ones for the identification of this K\"ahler structure. 
The complex fields are indeed easily identified as
\be
& S=e^{3\sigma} + i a_1 \, ,& \nonumber\\
&T=e^{\gamma} + i B,&
\label{moduliSU(3)}
\ee
and the corresponding K\"ahler potential is ${\cal K}=-\ln (S+S^{\dagger}) 
-3 \ln (T+T^{\dagger})$.
The imaginary part of $S$ is obtained from a Hodge duality of $G$
\be
e^{6\sigma} G_{\mu\nu\rho 5} = \epsilon_{\mu\nu\rho\sigma} \dhsigma a_1 \ .
\ee
Notice, first of all, the well-known exchange of roles $S-T$ compared
to string compactifications.
Moreover, unlike in the direct compactification of ten-dimensional 
string case, 
the definitions of $T$ and $S$ are completely decoupled, $S$ being 
related
to the volume of the 6-d compactified space and $T$ to the radius of the
eleventh dimension, but seen from 5-d. We call it the radius of the
fifth dimension $R_5$ in the following, in $M$-theory units (denoted 
by $R_5^{(M)}$ in section 4).  
We shall see in the last section, how, with another interpretation of 
this radius, we can reproduce the K\"ahler structure of usual 
string compactifications.

The $SU(3)$ invariance requirement on the CY manifold allows for one single 
field $\sigma$ in the metric.
We can in fact mimicke compactification on orbifolds by restriction to 
discrete subgroups of $SU(3)$ that act non-trivially
on the representation $3$ and ${\bar 3}$ in such a way that we maintain 
$N=2$ in five dimensions, by using the methods employed in 
\cite{CYasymmetric}.
We shall consider three cases (the action is on the representation $3$):
\begin{enumerate}
\item {\it $Z_{12}$ symmetry} acting like 
$(ie^{2i\pi /3},-ie^{2i\pi /3},e^{2i\pi /3})$ ;
\item {\it $Z_3$ symmetry} acting like 
$(e^{2i\pi /3},e^{2i\pi /3},e^{2i\pi /3})$  ;
\item {\it $Z_2\times Z_2^\prime$ symmetry} acting like 
$(-1,1,-1)\times (1,-1,-1)$ .
\end{enumerate}

{\bf a.} {\it $Z_{12}$ symmetry}. This model has a massless spectrum in 5-d
consisting of the universal hypermultiplet, $2$ vector multiplets and
the gravitational multiplet, corresponding to $h_{(1,1)}=3, h_{(2,1)}=0$.
The observable gauge group is $E_6\times U(1) \times U(1)$.
In the supergravity coordinates, the metric tensor takes then the form
(written directly from 11-d $\to$ 4-d)
\be
g_{\mu\nu}^{(11)}=e^{-\gamma-2(\sigma_1+\sigma_2+\sigma_3)/3} 
g_{\mu\nu}^{(4)},\
g_{55}^{(11)}=e^{2\gamma-2(\sigma_1+\sigma_2+\sigma_3)/3},\
g_{i {\bar \jmath}}^{(11)} = e^{\sigma_i} \delta_{i {\bar \jmath}},\
\ee
where the shape of the orbifold is here described by the three radii  
$\sigma_1,\sigma_2$ and $\sigma_3$.
 After the $Z_2^{HW}$ projection, the massless modes of the three form 
in 4-d are
\be
C_{\mu\nu 5} ,\ C_{5 i {\bar \jmath}} = B_{i} \delta_{i {\bar \jmath}}.
\ee
It is easy to perform the dimensional reduction of $S^{(5)}$ and, 
as previously, 
identify the K\"ahler structure.
The 4-d action now derives from 
\be
{\cal K} = -\ln (S+S^\dagger) - \sum_{1\leq k \leq 3} \ln (T_{k}+{T_{k}}^\dagger),
\ee
and the complex fields are identified as
\be
S &=& e^{\sigma_1+\sigma_2+\sigma_3} + i \, a_1 \ , \nonumber\\
T_{k} &=& e^{\gamma} e^{-(\sigma_1+\sigma_2+\sigma_3)/3+\sigma_k} + 
i\,  B_{k},
\ee
where $a_1$ is still defined by a Hodge duality
$ e^{2(\sigma_1+\sigma_2+\sigma_3)} G_{\mu\nu\rho 5}  = \epsilon_{\mu\nu\rho\sigma} 
\dhsigma a_1 \ .
$
Notice here the relations $t_i / t_j = R_i^2 / R_j^2$, 
$t_1 t_2 t_3 = R_5^3$, $s=R_1^2 R_2^2 R_3^2$, where $R_i$ denote the three 
CY radii. An interesting fact is that under
a dilatation of the 6-d compactified space $g_{i {\bar \jmath}}
\rightarrow  \lambda g_{i {\bar \jmath}}$, the only field which changes 
is $S$.
This will be valid for the more general truncations discussed below and
is related to the special geometry structure in 5-d.

\vspace{.2cm}
{\bf b.} {\it $Z_{3}$ symmetry}. 
The massless fields in 5-d  are the gravitational multiplet, the universal
hypermultiplet and eight vectors multiplets, corresponding
to $h_{(1,1)}=9$ and $h_{(2,1)}=0$ and the observable gauge group
is $E_6\times SU(3)$.
In the supergravity coordinates, the $Z_{3}$-invariant metric tensor 
takes the form
\be
g_{\mu\nu}^{(11)}=e^{-\gamma} {G}^{-1/3} g_{\mu\nu}^{(4)},\
g_{55}^{(11)}=e^{2\gamma} {G}^{-1/3},\
g_{i {\bar \jmath}}^{(11)} = {g}_{i {\bar \jmath}} \ .
\ee
The shape of the orbifold is here described by the nine parameters 
appearing in the
 unrestricted ${g}_{i {\bar \jmath}}$.
The massless modes of the three form in 4-d are
$  C_{\mu\nu 5} ,\ C_{5 i {\bar \jmath}} = B_{i{\bar \jmath}}\ .$
The resulting compactified 4-d action derives now from
 the following K\"ahler potential
\be
{\cal K} = -\ln (S+S^\dagger) - \ln \det (T_{i {\bar \jmath}}+
{T_{i{\bar \jmath}}}^\dagger),
\ee
and the complex fields are identified as
\be
S &=& G^{1/2}+ i\ a_1 \ ,\nonumber\\
T_{i{\bar \jmath}} &=& e^{\gamma} G^{-1/6}\, {g}_{i {\bar \jmath}} + i\ 
B_{i{\bar \jmath}}\ ,
\ee
where
$ G^2 G_{\mu\nu\rho 5} = \epsilon_{\mu\nu\rho\sigma} \dhsigma a_1 \ .$
Notice the relation $\det t_{i {\bar \jmath}} =e^{3\gamma}=R_5^3$ which 
is again related to special geometry in 5-d.

\vspace{.2cm}
{\bf c.} {\it $Z_{2}\times Z_2^{\prime}$ symmetry}.
Here, in contrast with previous cases, we obtain three additional 
hypermultiplets (in addition to the universal one), containing the bosonic
fields $(g_{ii}, C_{i {\bar \jmath} {\bar k}})$. The model is described by
$h_{(1,1)}=3, h_{(2,1)}=3$ and the observable gauge group on the boundary is 
$E_6\times U(1) \times U(1)$.  
In the supergravity coordinates, the $Z_{2}\times Z_2^{\prime}$-invariant
 metric tensor takes now the form
\be
g_{\mu\nu}^{(11)}=e^{-\gamma} {G}^{-1/3} g_{\mu\nu}^{(4)},\
g_{55}^{(11)}=e^{2\gamma} {G}^{-1/3}\
\ee
and
\be
g^{(11)}_{ij}= 
\left(
\begin{array}{ccc}
g^{(1)}&&\\
& g^{(2)}&\\
&& g^{(3)}
\end{array}
\right)
\ , 
\ee
where $g^{(i)}$ are $2 \times 2$ symmetric matrices, $G_i$ their 
determinant and $G$ the global determinant.

\noindent The 4-d massless modes of the three form are
$C_{\mu\nu 5} , \ C_{5 i {\bar \jmath}} = 
B_{i}\delta_{i {\bar \jmath}}\ .$
The corresponding 4-d K\"ahler potential is now
\be
{\cal K} = -\ln (S+S^\dagger) - \sum_{1\leq k \leq 3} 
\ln (T_{k}+{T_{k}}^\dagger) - \sum_{1\leq k \leq 3} 
\ln (U_{k}+{U_{k}}^\dagger),
\ee
where 
\be
S &=& G^{1/2} + i\  a_1 \, ,\nonumber\\
T_{k} &=& e^{\gamma} G^{-1/6} G_k^{1/2} + i B_{k}\ , 
\label{moduliZ2Z2}\\
U_{k} &=& {{(G_k)^{1/2}}\over {g^{(k)}_{22}}} + i\ 
{{g^{(k)}_{12}} \over{g^{(k)}_{22}}}\ ,\nonumber
\ee
with
$ G\,  G_{\mu\nu\rho 5}  = \epsilon_{\mu\nu\rho\sigma} \dhsigma a_1 \ .$

Notice that all the models studied above are particular cases of
no-scale models \cite{cremmer}. So the no-scale structure seems to
be present in both the weak-coupling limit and strong-coupling
limit of superstring compactifications \cite{phenoMtheory},\cite{li}.

Recently, it was shown that the anomaly cancellation in $M$-theory ask for  
another term in
the $M$-theory Lagrangian \cite{Horava&witten}, \cite{da}, which can be
 viewed as a term
cancelling fivebrane worldvolume anomalies \cite{dlm} or, by
compactifying one dimension, as a one-loop term in the type $IIA$ superstring 
\cite{vw}. In our conventions and by using differential form language, it 
reads
\be
W_5 = -{1 \over 2^{11/6} (2 \pi )^{10/3} } \int_{M_{11}} 
C \wedge X_8 \ ,
\label{eq:ar014}
\ee
where $X_8 \equiv -{1 \over 8} tr R^4 + {1 \over 32} (tr R^2)^2$.
This term can be compactified to 4-d in different compactification
schemes by using the field definitions given above and the quantization
rule ${1 \over 8 \pi^2} \int tr R \wedge R=m$,
where $m$ is an integer and the integration is over a $4$-cycle in the
compactified space. Without entering into details and restricting to one
overall radius, we find in 4-d a higher-derivative term of the type $T R^2$ 
in 
superfield language (recently, similar 4-d terms were studied \cite{ovrut}
and claimed to be relevant for supersymmetry breaking).

  Armed with the above field definitions, we can now study our main concern,
supersymmetry breaking  from $5 \rightarrow 4$ dimensions by compactification.


\section{The spontaneous breaking of $N=1$ supersymmetry in four dimensions 
by compactification}

We are interested in the phenomenology of the $M$-theory compactifications,
in particular we would like to break the $N=1$ supersymmetry in 4-d. 
To achieve 
this aim, we perform a mechanism proposed by Scherk and Schwarz 
\cite{Scherk&Schwarz}
in a supergravity context and then generalized to superstrings in 
\cite{kounnas}
(another possibility for breaking supersymmetry is the gaugino 
condensation of one of the 
$E_8$ gauge group \cite{horava}).
As for the moment there is no quantum description of
$M$-theory, we use its effective low energy description at the level of
supergravity.

The Scherk-Schwarz mechanism is a generalized dimensional reduction which 
allows for the 
fields a dependence in the compact coordinates. Nevertheless, to avoid 
ghost particles and to include ordinary dimensional reduction, this 
dependence must satisfy some properties: 
it has to be in a factorisable form and has to correspond to a 
R-symmetry of the theory. The simplest example is the use of 
the compact $SO(6)$ symmetry of the
6-d compact manifold, which is readily applicable to the 
superstrings case.  In this case, any component of a tensor 
with $p$ compact indices and $q$ non-compact indices takes the form
\be
{\hat T}_{i_1 \cdots i_p {\mu}_1 \cdots {\mu}_q} (x,y)
&=& \prod_{n=1}^{p} U_{i_n}^{i'_n} (y) \
{ T}_{i'_1 \cdots i'_n {\mu}_1 \cdots {\mu}_q} (x) ,
\ee
where $y$ denotes compact coordinates and $x$ non compact ones.
This tensor decomposition is stable under product and exterior derivation.

This extended dimensional reduction, when applied to the curvature term, 
generates a potential for the scalar fields corresponding to the metric 
in the compact space. The requirement for this scalar potential to be 
positive imposes further restrictions on
 the form of $U$.
A solution was proposed by Scherk and Schwarz, by taking
\be
U &=& e^{M y^1}\ ,
\label{defUM}
\ee
where $M$ is an antisymmetric matrix in the compact space with zeros in 
the row and the column corresponding to $y^1$.
Then the 4-d scalar potential, in supergravity units, reads \cite{Scherk&Schwarz}
\be
V_0 &=&
{1\over 4} {1\over \sqrt{G}}
\left( g^{1,1} tr \left( M^2 - M g M g^{-1} \right)
- \left( g^{-1} M^t g M g^{-1} \right)^{1,1}
\right),
\label{V0}
\ee
where in (\ref{V0}) and the rest of this section, $g$ is the metric 
in the 7-d compact space and $G$ its determinant.

When applied to the kinetic term for the 11-d spin-${3\over 2}$ field, 
this extended dimensional reduction also generates, through the spin 
connection for compact indices, masses for the resulting 4-d gravitini.
 
More important for our purposes are symmetries which mix fields of 
different
parities, for reasons which will become transparent below. For the 
truncations we are
discussing here, these must be subgroups of the $Sp(8)$ symmetry present
in the $N=8$ supergravity in 5-d ( see third reference in 
\cite{Scherk&Schwarz}).

We shall perform this mechanism in two cases
\begin{enumerate}
\vspace{-2mm}
\item with $y^1$ being the extra eleventh dimension, $x^5$;
\item with $y^1$ being a Calabi-Yau or orbifold coordinate .
\end{enumerate}
 We use for this purpose the truncations derived in the preceding section. 
The projections $P$ we use have to commute with the Scherk-Schwarz 
matrix $U$,
which means explicitly
\be
&[ P, M] = 0 \ \ if \ \ P y^1 = y^1 \ , &\nonumber\\
&\{ P,M \}= 0  \ \ if \ \ P y^1 = -y^1 \ .&
\label{commuGU}
\ee

{\bf a.} {\it An eleventh dimension coordinate dependence}.
The characteristics of this case is the anticommutation relation
\be
\{ Z_2^{HW} , M \} = 0 \ . \label{eq:s0}
\ee

We consider here two different possibilities. The first uses an $SU(2)$
R-symmetry present in the 5-d theory, which we argue to be the subgroup
of $Sp(8)$ which commutes with all the projections. The second uses the
usual symmetry of the 6-d internal manifold, which is relevant for the
type $IIA$ supergravity in 4-d.

{\bf i.}  $SU(2)$ {\it R-symmetry}.

To keep things as simple as possible, consider for the moment the
simplest truncation (\ref{metricto5d}) corresponding to an $SU(3)$ invariance
in the compactified space. In this case, in 5-d the only matter
multiplet is the universal hypermultiplet, whose scalar fields parametrize
the coset ${{SU(2,1)} \over {SU(2) \times U(1)}}$ \cite{ferrara}. 
This structure can be simply viewed from 4-d by a direct truncation 
({\it without} the $Z_2^{HW}$ projection). The truncated dependence on the
universal hypermultiplet can be derived from the K\"ahler potential 
\cite{ferrara}
\be
{\cal K} =- \ln (S+S^{\dagger} - 2 a^{\dagger} a) \ , \label{eq:s1}
\ee
where the Hodge duality in 5-d is 
$G_{\mu\nu\rho\sigma} = \epsilon_{\mu\nu\rho\sigma\delta} 
( -\partial^{\delta} a_1 + i a^\dagger \partialleftright a / \sqrt{2})$
and
$S=e^{3 \sigma}+a^{\dagger} a + i a_1$. The $SU(2)$
symmetry acts linearly on the redefined fields
\be
z_1 = {1-S \over 1+S} \ , \ z_2 = {2 a \over 1+S} \ , \label{eq:s2}
\ee
which form a doublet $(z_1, z_2)$. The $Z_2^{HW}$ projection acts as
$Z_2^{HW} S =S$, $Z_2^{HW} a = -a$, which translates on the $SU(2)$ doublet
in the obvious way
\be 
Z_2^{HW}
\left( 
\begin{array}{c} 
z_1 \\ 
z_2 
\end{array} 
\right)  
&=& 
\left(
\begin{array}{cc} 
1 & 0 \\ 
0 & -1
\end{array}
\right)
\left(
\begin{array}{c}
z_1 \\
z_2
\end{array}
\right) \ . \label{eq:s3}
\ee

The Scherk-Schwarz decomposition in this case reads explicitly
\be 
\left( 
\begin{array}{c} 
{\hat z}_1 \\ 
{\hat z}_2 
\end{array} 
\right)  
&=& 
\left(
\begin{array}{cc} 
\cos m x_5 & \sin m x_5 \\ 
-\sin m x_5 & \cos m x_5
\end{array}
\right)
\left(
\begin{array}{c}
z_1 \\
z_2
\end{array}
\right) \ ,\label{eq:s4}
\ee
corresponding to the matrix defined in (\ref{defUM}) $M=im \sigma_2$ and 
where $m$ is a mass parameter. 
Notice that, 
thanks to the anticommutation relation (\ref{eq:s0}) which is clearly verified, 
the fields ${\hat z}_i$ have
the same $Z_2^{HW}$ parities as the fields $z_i$.
The resulting scalar potential in 4-d in Einstein metric is computed from the 
kinetic terms of the $({\hat z}_1, {\hat z}_2)$ fields derived form (\ref{eq:s1}). 
After 
putting $z_2=0$ corresponding to the projection $Z_2^{HW}$, it is
 easily worked out 
and it is
positive semi-definite. Expressed in terms of $S$ and $T$,
the result is
\be
V = {4 m^2 \over (T+T^{\dagger})^3} {|1-S|^2 \over S+S^{\dagger}}
\ . \label{eq:s40}
\ee

This result can be seen as a superpotential generation for $S$. The 4-d
theory is completely described 
by\footnote{This result is in agreement with general results on no-scale 
models (second ref. in \cite{kounnas}).}
\be
&{\cal K} =-\ln (S+S^{\dagger}) - 3 \ln (T+T^{\dagger}) \ ,& \nonumber  \\
&W= m (1 + S) \ .& \label{eq:s5}
\ee
Notice that this superpotential for $S$ should correspond to a
non-perturbative effect from the heterotic string point of view.
The minimum of the scalar potential corresponds to $S=1$ and a 
spontaneously
broken supergravity model with a zero
cosmological constant. The order parameter for supersymmetry breaking
is the gravitino mass $m_{3/2}^{2} =e^{{\cal K}} |W|^2 = 
{2 m^2 / (T+T^{\dagger})^3}$.
Notice that the theory (\ref{eq:s5}) is symmetric under the inversion
$S \rightarrow 1/S$, easily checked also on the scalar potential
 (\ref{eq:s40}).
This corresponds to a subgroup of the $SL(2,Z)$ acting on S as
$S \rightarrow (aS-ib)/(icS+d)$, where $ad-bc=1$, restricted here to
$a=d=0$, $b=-c=1$. This already suggest a weak coupling - strong coupling
symmetry of our resulting model, as in the $S$-duality models proposed some
time ago \cite{ibanez}.

 We stress out that this 4-d model presents some general features, 
independent
of the details of the compactification\footnote{The following 
arguments were developed in collaboration with R. Minasian.}.
The most general R-symmetry present in 5-d in N=8 supergravity is $Sp(8)$.
This symmetry is reduced once we impose truncations, and the truncation to
N=2 leave an $SU(2)$ symmetry appearing in the supersymmetry algebra. 
The scalar fields 
from the universal hypermultiplet, present in any compactification, 
transform as 
a doublet under $SU(2)$. Scherk-Schwarz 
mechanism uses only
the antisymmetric part ($O(2)$) of it, so we are led to (\ref{eq:s4}).
On the other hand, the $O(2)$ part is exactly which is required by the
Horava-Witten $Z_2^{HW}$ projection which defines the $E_8 \times E_8$
heterotic string. In more general CY models in 5-d,
it is maybe possible that the $SU(2)$ symmetry acts also on other scalar 
fields than the universal hypermultiplet, but the piece (\ref{eq:s5})
we computed should always exist and is universal.

{\bf ii.} {\it Symmetry of the compactified space}.

Since $P$ ($SU(3)$, $Z_3$,\, $Z_{12}$ or $Z_2 \times Z'_2$) doesn't act on $x^5$, 
the commutation relation (\ref{commuGU}) simply reads
\be
[ P, M ] = 0 .
\ee
For $P=SU(3)$, there is no antisymmetric matrix $M$ respecting this
 condition, 
so we can't perform the 
Scherk-Schwarz's mechanism.
For $P=Z_2\times Z_2^\prime$, the allowed M matrices involve three 
parameters 
and, in the basis $(x^5,x^6,\cdots , x^{11})$, read
\be
M= \left(
\begin{array}{ccccccc}
0&&&\cdots&&&0\\
&0&m_1\\
&-m_1&0\\
\vdots&&&0&m_2\\
&&&-m_2&0\\
&&&&&0&m_3\\
0&&&&&-m_3&0
\end{array}
\right).
\ee
In this case, the Scherk-Schwarz matrix doesn't anticommute with $Z_2^{HW}$,
so the mechanism is better adapted to type $IIA$ supergravity in 4-d.
We are only interested here in field spectrum common to type $IIA$ and 
heterotic string.
As we will see in the next section, the qualitative features of this case  
concerning the decompactification problem are however similar to that
of the previous example.

The scalar potential (\ref{V0}) in this case reads
\be
V_0 & = & 
{1\over 4} e^{-3 \gamma} \sum_{i=1}^{3}
{{m_{i}^2}\over{G_i}} \left( \left( g_{11}^{(i)}-g_{22}^{(i)} \right)^2 + 
4 {g_{12}^{(i)}}^2 \right) ,
\ee
which, expressed in terms of moduli fields (\ref{moduliZ2Z2}), takes the 
simple expression
\be
V_0 & = & 
{1 \over {t_1 t_2 t_3 }} \sum_{i=1}^{3}
m_i^2 \left| {{U_i^2-1}\over {U_i+U_i^\dagger}} \right|^2.
\ee
This scalar potential has interesting consequences. In analogy with our
previous example, the scalar potential is invariant under $U_i \rightarrow
1/U_i$. The three $U$-fields 
develop a {\it vev} $< U_i > =  1$ (the self-dual points of duality
transformations) and acquire masses
\be
m_{U_i}^2 & = & {{ m_i^2}\over {<t_1 t_2 t_3>}}.
\ee
At the same time, the single gravitino which remains after imposing the 
$Z_2\times Z_2^\prime\times Z_2^{HW}$ symmetries, also acquires a mass
 which is computed to be
\be
m_{3/2} & =& {{m_1+m_2+m_3}\over{2 \sqrt{<t_1 t_2 t_3>} }} \ .
\label{gravitinomassx^5}
\ee

{\bf b.} {\it A Calabi-Yau or orbifold coordinate dependence}.
We choose here $y^1$ to be $x^6$, a Calabi-Yau or orbifold coordinate. 
Then the 
commutation relation (\ref{commuGU}) reads
\be
[ P, M x^6 ] = 0 \ , \ [Z_2^{HW} , M x^6] = 0 \ .
\label{commuGM}
\ee
For $P=SU(3)$, there is no antisymmetric matrix $M$ respecting this 
condition, 
so we can't perform the 
Scherk-Schwarz's mechanism.
For $P=Z_2\times Z_2^\prime$, the situation is different; since $x^6$ is 
now odd under 
the first $Z_2$ and even under the second one and under $Z_2^{HW}$, the 
commutation 
relation simplifies again
\be
\{ Z_2 , M \} = 0 , \ \ [Z_2^\prime , M ] =0, \ \ [Z_2^{HW} , M] = 0 \ . 
\ee
Thus we can always choose $M$ to be of the off-diagonal following form, 
in the basis 
$(x^5,x^6,\cdots,x^{11})$,
\be
M= \left(
\begin{array}{ccccccc}
\\
&&&&0&&\\
\\
&&0&&&m&0\\
&&&&&0&m^\prime\\
&&&-m&0\\
&&&0&-m^{\prime}
\end{array}
\right) \ . \label{eq:s7}
\ee
Then the scalar potential, written directly in term of moduli fields, 
reads
\be
V_0 & = & {1\over{4st_1u_1t_2u_2t_3u_3}} \left(
m^2 |t_2U_2-t_3U_3 |^2 + {m^{\prime}}^2 |t_2U_3-t_3U_2|^2 \right.\\
&& \left. 
-{1 \over 2} \left( (m-m^\prime)^2t_2t_3 - mm^\prime (t_2-t_3)^2 \right) 
(U_2 -{U^{\dagger}}_2 ) (U_3 - {U^{\dagger}}_3) \right) \ .
\nonumber
\ee
The minimum (for $m \not = m'$)  of the scalar potential corresponds to
the flat directions
\be
<t_2>=<t_3>,\ <u_2>=<u_3> \ \mbox{and} <\Imm  U_2> = <\Imm U_3> = 0.
\ee  
We don't write here the scalar mass matrix, but just notice that the
leading matrix elements, which fix the physical masses, are of the
type (consider $m \sim m^\prime$)
\be
m_0^2 \sim {m^2 \over s t_1} \ , \label{eq:s70}
\ee 
where $t_1$ is the modulus field
corresponding to the Scherk-Schwarz coordinate. This case is the analog of 
the supersymmetry breaking by compactification in the weakly 
coupled superstrings \cite{kounnas}. 
Notice that in the superstring case, the mass parameters $m,m'$ in 
(\ref{eq:s7}) are discrete, being related to automorphisms of the
compactification lattice.

All the masses we computed above are measured in $M$-theory  units. 
In the 
following section, 
we translate these results in 4-d supergravity units.

\section{Phenomenological implications of $M$-theory compactifications}

We shall discuss in this section some phenomenological isssues of $M$-theory 
compactifications 
focussing essentially on the four dimensional Newton's constant and the 
gravitino mass. 
For simplicity reasons and without affecting our conclusions, we restrict 
ourselves to
the simplest case $t_i=e^\gamma$.

String compactifications don't easily allow \cite{stringscale85},
\cite{stringdualities} to adjust
 the string mass scale, the {\it vev } of the dilaton and the volume of 
the compact space to tune
 the three four-dimensional observable quantities that are the 
Planck scale, the GUT scale and 
the gauge coupling constant. For instance, in the weakly coupled
 regime of the $E_8\times E_8$ heterotic string, the well-known 
relation 
\be
G_{{\cal N}} \sim \lambda_{st}^{2/3} 
{{\alpha_{GUT}^{4/3}}\over{M_{GUT}^{2}}}
\label{relweaklycoupled}
\ee
disagrees with experimental values by a factor of order $20$.
Witten has shown \cite{stringdualities} how in strongly coupled regime 
the compatibility 
between string predictions and experimental values could
be restored. It is interesting to worry about this issue in our patterns 
of compactification.

In the 11-d action appears only the $M$-theory scale, which is the 
eleven-dimensional 
Planck mass $M_{11}$, such that
\be
S^{(11)} \supset
{-1\over 2}\,
\int d^{11}x \sqrt{g^{(11)}} M_{11}^9 {\cal R}^{(11)}
- {1 \over {8 \pi (4 \pi )^{2/3}}} \int_{x^5=0} d^{10}x 
\sqrt{g^{(10)}} M_{11}^6 F_{\mu\nu} F^{\mu\nu}.
\ee
Now we compactify this action in the $M$-theory coordinates in which, like 
in the five-brane units for the 10-d string, there is no kinetic term for 
the "radius" of the extra-dimension, that is we adopt
\be
& g_{\mu\nu}^{(11)} = e^{-2\sigma} g_{\mu\nu}^{(5)} , \ 
g_{i {\bar \jmath}}^{(11)} = e^{\sigma} {\delta}_{i {\bar \jmath}} 
\,  , & \nonumber \\
& g_{55}^{(5)} = e^{2\gamma} , \
g_{\mu \nu}^{(5)} =  g_{\mu \nu}^{(4)} \ .&  
\label{$M$-theorycoordinates}
\ee
It is transparent that these are the natural units of 5-d supergravity.
By performing the Hodge transformation (using the field definitions
(\ref{moduliSU(3)}))
\be
{s \over t} \ G_{\mu \nu \rho 5} &=&
\epsilon_{\mu \nu \rho \sigma} \partial^{\sigma} a^\prime\ ,
\ee
the 4-d low energy effective action contain the terms
\be
S^{(4)} \supset
{-1\over 2} \,\int d^4x \sqrt{g^{(4)}} 
\left[ t M_{11}^2 
\left( {\cal R}^{(4)} 
+{1 \over 2 s^2 } ( {\partial_{\mu} s})^2 + ({\partial_{\mu} a^\prime})^2
\right) 
+{1 \over { (4 \pi )^{5/3}}} s\, F_{\mu\nu} F^{\mu\nu} 
\right] \ .
\label{eq:s30} 
\ee
We can pass from the $M$-theory action (\ref{eq:s30}) to the Einstein action
(\ref{action4-d}) by the Weyl rescaling $g_{E}^{(4)}=t\, g_{M}^{(4)}$.
Notice the absence of a kinetic term for $t$ in (\ref{eq:s30}), the clear
indication of a no-scale structure. This can therefore be associated with
the $M$-theory (or 5-d supergravity) units.

The Lagrangian (\ref{eq:s30}) allows us to identify the gauge coupling and the 
4-d reduced Planck mass to be \footnote{In weakly-coupled regime, the gauge
coupling constant can have big threshold corrections. If this holds true in 
strongly coupled regime, some
of our considerations below will be modified.}
\be
{1 \over \alpha_{GUT}} \sim {1 \over {(4 \pi )^{5/3}}}
s, \ \ t M_{11}^2 = {M_{Pl}^{(4)}}^{2} \ . \label{eq:s31}
\ee
These quanitities are related to the radii of the Calabi-Yau space and the
extra  fifth dimension expressed in terms of the $M$-theory units
\be
e^{-\sigma/2} =s^{1/6}= {M_{GUT} \over M_{11}}, \ \ t = R_5^{(M)} M_{11}.
\ee
However the radius of the fifth dimension expressed in $M$-theory units is not convenient 
for physics in four dimensions, so we prefer to express it in terms
of $M_{Pl}^{(4)}$, the four-dimensional Planck scale, by defining
\be
t = {M_{Pl}^{(4)} \over R_5^{-1}} \ .
\label{defM_5}
\ee
Combining these equalities, we obtain the $M$-theory equivalent of (\ref{relweaklycoupled})
\be
G_{{\cal N}} \sim {{R_5^{-1} \alpha_{GUT}^{1/3}}\over{M_{Pl}^{(4)}
M_{GUT}^2}} \ ,
\ee
and, in order to be compatible with experimental constraints, the "radius" of the fifth dimension 
has an intermediate scale value, as already suggested by 
\cite{phenoMtheory},\cite{aq}
\be
R_5^{-1} \sim 10^{12-13} GeV.
\ee

The fifth radius also appears in the gravitino mass generated by the Scherk-Schwarz mechanism, 
so we have to take care that $R_5^{-1}$ is compatible with 
phenomenological prefered value for 
the gravitino mass.
In the expressions for the gravitino mass from (\ref{eq:s5}) and 
(\ref{gravitinomassx^5}), the mass parameters that 
appear in the Scherk-Schwarz mechanism using the fifth dimension are 
naturally of the order of the 
$M$-theory scale, 
so we obtain\footnote{A similar formula was
conjectured by Antoniadis and Quiros \cite{aq}, but replacing $R_5$ with
the eleventh radius. We stress out that in our expression, $R_5$ is the radius
of the eleventh dimension seen from 5-d in the 5-d supergravity (Einstein)
units.}
\be
m_{3/2} \sim {{M_{11}}\over {\sqrt{<t^3>}}} = {R_5^{-2} \over M_{Pl}^{(4)} } 
\ ,
\ee
where to obtain the last identity we have used the identification of the 
moduli fields 
(\ref{moduliSU(3)}) and the definitions (\ref{eq:s31}), (\ref{defM_5}).
This expression is similar to a usual supergravity form in which an 
auxilary field 
would have developped a {\it vev}
of the order of $R_5^{-2}$.
Phenomenologically, the gravitino mass should be of the order of 
$1\, TeV$. 
This leads to a value 
for the extra-radius rather compatible with the one deduced from Newton's constant,
\be
R_5^{-1} \sim 10^{11-12} GeV.
\ee
Consequently, the presence of the extra-dimension could offer a solution to 
the 
decompactification 
puzzle usually encountered in string compactifications.
Indeed generically, Scherk-Schwarz mechanism in strings leads to 
\cite{kounnas}
\be
m_{3/2} \sim {1\over {R_{CY}}} \sim M_{GUT}.
\ee
This is actually similar to the result we got in the last section in 
the case where a CY
coordinate was used for the Scherk-Schwarz mechanism. To check it, rewrite
the masses (\ref{eq:s70}) in 4-d supergravity units  as 
$m_0^2 \sim {M_{Pl}^{(4)}}^{2} / (s t^2)
\sim g^2/R_5^2$, where $g$ is the gauge coupling.

This relation is phenomenologically hard to accomodate. Two issues 
have been proposed to 
solve this problem:

- compactify six dimensions on an asymmetric Calabi-Yau space with two or
more different radii \cite{amq}. However to agree with phenomenological 
values for 
the gravitino mass,
these two (or more) radii must have hierarchical values by about fifteen 
orders of magnitude.

- construct models of strings such that the gravitino mass is 
$m_{3/2} \sim 1/(R_{CY}^{n+1} {M_{Pl}^{(4)}}^{n}) \sim (M_{GUT} / M_{Pl}^{(4)})^n
M_{GUT}$. However, phenomenology asks for $n=4,\, 5$, and is difficult to 
explicitly construct such models.

So the extra-dimension brings a new and more satisfying alternative to the 
decompactification 
problem of compactified strings.

To conclude this phenomenological study, we would like to compare our 
pattern of 
compactification $(11\to 5 \to 4)$ with another one already studied
\cite{Dudas&Mourad} in which the extra-dimension was compactified first
$(11\to 10 \to 4)$.
From a strongly coupled 10-d point of view, the radius of the eleventh 
dimension $x^5$ is
$ R_{11}^2 \sim g_{55}^{(11)} \sim e^{2\gamma-2\sigma},$
and this radius is related to the string dilaton by \cite{stringdualities}
\be
R_{11} \sim e^{2\phi /3}.
\label{R11-dilaton}
\ee
Combining those relations, the expressions of the real part of $T$ becomes
$ t \sim e^{2\phi /3} e^\sigma \, ,$
which is the expression obtained by Bin\'etruy  \cite{Binetruy} by 
compactifying 
the 10-d heterotic string in the five-brane units which are supposed to be 
appropriate to
the strongly coupled regime.

Actually we can, in the same way, reproduce the moduli's identification
obtained 
by string compactification down to 4-d in the various units.
Suppose that we are working in units caracterized, in 10-d, by the 
tensor metric 
$g_{{\hat \mu}{\hat \nu}}^{(10)}$.
These units are related to the five-brane ones by a Weyl rescaling
\be
g_{{\hat \mu}{\hat \nu}}^{(10)} &=& \lambda \ g_{{\hat \mu}{\hat \nu}}^{(10)\ B}.
\label{five-braneunits}
\ee
So the typical radius of the Calabi-Yau space, measured in five-brane units, 
is now 
$R_{CY} \sim (\lambda^{-1} e^{\sigma})^{1/2} \equiv (e^{\sigma_B})^{1/2}$
and the radius of the extra-dimension, viewed from 10-d in five-brane units, 
is
$R_{11} \sim e^{\gamma-\sigma_B}$
which is still related to string dilaton by (\ref{R11-dilaton}).
\noindent The real parts of the fields are now identified as
\be
s & = & e^{3\sigma_B} \ = \ \lambda^{-3} e^{3 \sigma} \nonumber \\
t & = & e^{\gamma} \ = \ \lambda^{-1} e^{\sigma+2\phi/3}.
\label{modulivariousunits}
\ee
For instance, string (S) and supergravity (E) units are related 
to five-brane (B) units by
\be
g_{{\hat \mu}{\hat \nu}}^{(10)\ B} \ = 
\ e^{-2\phi/3} g_{{\hat \mu}{\hat \nu}}^{(10)\ S} 
\ = \ e^{-\phi/6} g_{{\hat \mu}{\hat \nu}}^{(10)\ E} \ .
\ee
Eqs. (\ref{modulivariousunits}) reproduce the results 
of string compactification 
summarized in tab.\ref{tab:moduli_in_various_units}.

\begin{table}[h]
\begin{center}
\begin{tabular}{cccc}
& five-brane units & string units & supergravity units \\
\hline
$\lambda$ \hspace{.2cm}& 1 & $e^{2\phi/3}$ & $e^{\phi/6}$ \\
$s$ & $e^{3\sigma}$ & $e^{3\sigma-2\phi}$ & $e^{3\sigma-\phi/2}$ \\
$t$ & $e^{\sigma+2\phi/3}$ & $e^{\sigma}$ & $e^{\sigma+\phi/2}$\\
\hline 
\end{tabular}
\end{center}
\caption{identification of the real parts of moduli fields in string 
compactification in
 various units \cite{Binetruy}.}
\label{tab:moduli_in_various_units}
\end{table}

However from a five-dimensional point of view, \ie after the 
compactification of the 
CY manifold, the radius of the eleventh dimension should be
$R_{11} \sim e^{\gamma}$,
so the identification of the moduli fields is now different, leading to 
the 
previous conclusions.

From eleven-dimensional point of view in $M$-theory coordinates 
(\ref{$M$-theorycoordinates}), the $11\to 5 \to 4$ pattern of 
compactification is consistent if $R_5 > R_{CY} $
which reads $t > s^{1/2}.$

From ten-dimensional point of view in five-brane coordinates 
(\ref{five-braneunits}), 
in the $11\to 10 \to 4$ pattern of compactification,
the radius of the Calabi-Yau space and of the extra-dimension are now
$ R_{CY} \sim e^{\sigma/2} \ \ \mbox{and} \ \ R_{11} \sim e^{2\phi/3} \ ,
$
while the consistency condition of compactification is $R_{CY} > R_{11}$, 
which, in term of moduli fields (tab.\ref{tab:moduli_in_various_units}), 
becomes now $t < s^{1/2}.$

In both cases, the ten-dimensional string strongly coupled regime condition 
$e^{\phi} >1$ simply reads $t > s^{1/3}.$
So we can draw, on the  $<t>$ line, the different consistent patterns of 
compactification (we assume that $s>1$)
$$\epsfbox{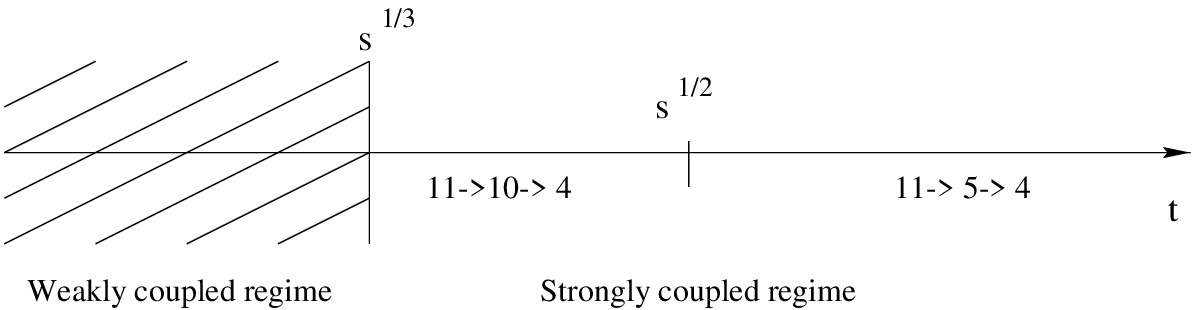}$$

We have seen that phenomenology asks for large values of $t$, so our
pattern of compactification is self-consistent.


\section{Conclusions}

We studied truncations of $M$-theory from $11 \rightarrow 5 \rightarrow 4$
dimensions, applying our results to the supersymmetry breaking
mechanism by compactification from 5-d to 4-d. The geometric properties
of the moduli fields, related to the special geometry of the 5-d theory are
peculiar and important for phenomenological purposes. $\Re e \,S$ is the 
volume
of 6-d internal manifold, the various $h_{1,1}$ moduli fields $T$ are 
related
to the radius of the fifth-dimension and are invariant under dilatations
of the 6-d manifold and the  $h_{2,1}$ moduli $U$  characterize, as 
usual,
the complex structure of the 6-d manifold. 

 Using different symmetries of the 5-d theory, we are able to break supersymmetry 
by compactification in various ways. The most interesting one
is by using an $SU(2)$ symmetry related to the universal hypermultiplet of
the 5-d theory. In this case, we obtain (for the simplest possible 
truncation)
a model described by

\be
&{\cal K} =-\ln (S+S^{\dagger}) - 3 \ln (T+T^{\dagger}) \ , &\nonumber  \\
&W= m (1 + S) \ .&
 \label{eq:s50}
\ee
This universal superpotential for $S$ should corresponds to a
non-perturbative string effect.
The minimization gives $S=1$ and a spontaneously
broken supergravity model with a zero
cosmological constant. We hope that this result will shed some
light on problems like dilaton stabilization and supersymmetry breaking
in effective strings. Another example, using symmetries of the compactified
6-d space gives rise to a potential for the
$U$ moduli, the corresponding vacuum  is $U_i =1$ with a zero cosmological constant.

Defining 4-d scales and couplings we find that the resulting gravitino mass
in supergravity units is $m_{3/2} \sim R_5^{-2} / M_{Pl}^{(4)}$, which
ameliorate the usual decompactification limit ($R_5 \rightarrow 0$)
problem and correlate it with the unification problem in a way which
look phenomenologically promising.

\section*{Acknowledgments}
It is a pleasure to thank P. Bin\'etruy, C. Kounnas, R. Minasian, J. Mourad, 
B. Pioline and C. Savoy for helpful discussions and comments.
  


\end{document}